\documentclass[sigconf]{acmart}

\AtBeginDocument{%
  }

\copyrightyear{2024}
\acmYear{2024}
\setcopyright{acmlicensed}
\acmConference[CIKM '24]{Proceedings of the 33rd ACM International Conference on Information and Knowledge Management}{October 21--25, 2024}{Boise, ID, USA}
\acmBooktitle{Proceedings of the 33rd ACM International Conference on Information and Knowledge Management (CIKM '24), October 21--25, 2024, Boise, ID, USA}
\acmDOI{10.1145/3627673.3679567}
\acmISBN{979-8-4007-0436-9/24/10}

\usepackage[ruled,vlined,linesnumbered]{algorithm2e}
\usepackage{multirow}
\usepackage{enumitem}
\usepackage{color}
\usepackage{balance}
\newcommand\modelName{MGHSTN}
\begin{document}

\title{Urban Traffic Accident Risk Prediction Revisited: Regionality, Proximity, Similarity and Sparsity}


\author{Minxiao Chen}
\orcid{0009-0004-0476-6174}
\affiliation{%
  \institution{Beijing University of Posts and Telecommunications}
  \city{Beijing}
  \country{China}
}
\affiliation{%
  \institution{Beiyou Shenzhen Institute}
  \city{Beijing}
  \country{China}
}
\email{chenminxiao@bupt.edu.cn}

\author{Haitao Yuan}
\authornotemark[1]
\orcid{0000-0001-6721-065X}
\affiliation{%
  \institution{Nanyang Technological University}
  \country{Singapore}
}
\email{haitao.yuan@ntu.edu.sg}

\author{Nan Jiang}
\orcid{0009-0001-3933-1662}
\affiliation{%
  \institution{Beijing University of Posts and Telecommunications}
  \city{Beijing}
  \country{China}
}
\affiliation{%
  \institution{Beiyou Shenzhen Institute}
  \city{Beijing}
  \country{China}
}
\email{jn_bupt@bupt.edu.cn}

\author{Zhifeng Bao}
\orcid{0000-0003-2477-381X}
\affiliation{%
 \institution{The Royal Melbourne Institute of Technology}
 \city{Melbourne}
 \country{Australia}}
\email{zhifeng.bao@rmit.edu.au}

\author{Shangguang Wang}
\orcid{0000-0001-7245-1298}
\affiliation{%
  \institution{Beiyou Shenzhen Institute}
  \city{Beijing}
  \country{China}
}
\affiliation{%
  \institution{Beijing University of Posts and Telecommunications}
  \city{Beijing}
  \country{China}
}
\email{sgwang@bupt.edu.cn}

\thanks{*Corresponding author.}
\renewcommand{\shortauthors}{Minxiao Chen, Haitao Yuan, Nan Jiang, Zhifeng Bao, \& Shangguang Wang}

\begin{abstract}
Traffic accidents pose a significant risk to human health and property safety. Therefore, to prevent traffic accidents, predicting their risks has garnered growing interest. We argue that a desired prediction solution should demonstrate resilience to the complexity of traffic accidents. In particular, it should adequately consider the regional background, accurately capture both spatial proximity and semantic similarity, and effectively address the sparsity of traffic accidents. However, these factors are often overlooked or difficult to incorporate. In this paper, we propose a novel multi-granularity hierarchical spatio-temporal network. Initially, we innovate by incorporating remote sensing data, facilitating the creation of hierarchical multi-granularity structure and the comprehension of regional background. We construct multiple high-level risk prediction tasks to enhance model's ability to cope with sparsity. Subsequently, to capture both spatial proximity and semantic similarity, region feature and multi-view graph undergo encoding processes to distill effective representations. Additionally, we propose message passing and adaptive temporal attention module that bridges different granularities and dynamically captures time correlations inherent in traffic accident patterns. At last, a multivariate hierarchical loss function is devised considering the complexity of the prediction purpose. Extensive experiments on two real datasets verify the superiority of our model against the state-of-the-art methods.
\end{abstract}




\maketitle

\section{INTRODUCTION}
According to the World Health Organization (WHO)~\cite{WHO}, approximately 1.3 million people die each year due to traffic accidents, making it a leading cause of death among children and young adults aged 5 to 29. In response, the United Nations General Assembly has set an ambitious target of halving the global number of deaths and injuries from road traffic crashes by 2030~\cite{UN}. Therefore, it is of vital importance to predict traffic accident risk accurately, which can assist government in managing traffic risks and help drivers avoid high-risk areas. When revisiting this problem, we find some critical yet missing aspects that contribute to its accuracy boost.

\begin{figure}
  \centering  
  \includegraphics[width=0.8\linewidth]{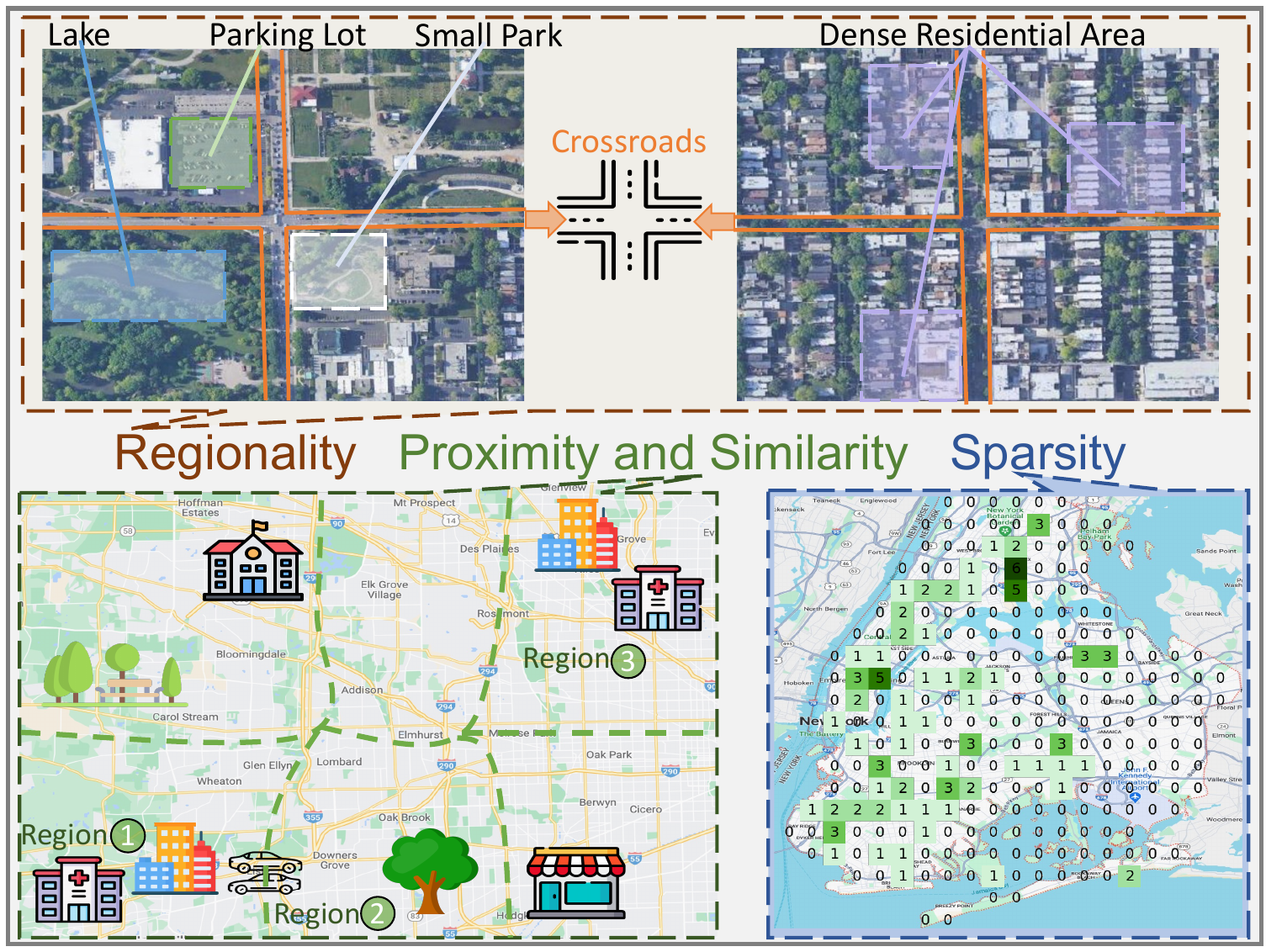}
  \vspace{-0.1in}
  \caption{Critical Aspects of Traffic Accident.}
  \vspace{-0.2in}
  \label{fig:challenges}
\end{figure}

\noindent (1) \textbf{Regionality: Regional Background Effects.} Traffic accidents are affected by many complex factors, and the ways to which these factors affect are difficult to analyze. To capture the underlying mechanisms of traffic accidents, regional background is essential. For instance, as shown in Fig.~\ref{fig:challenges}, different regions may have share similar road structure (e.g., crossroads) and weather (e.g., sunny), but entirely different regional background (e.g., parking lot, lake, small park vs. dense residential area). This leads to significant variations in traffic accident patterns. However, these regional backgrounds are often overlooked and difficult to capture.

\noindent (2) \textbf{Proximity and Similarity: Spatial and Semantic Correlations.} Traffic accidents exhibit dual correlation involving both spatial proximity and semantic similarity. This implies that traffic accidents occurring in close areas or areas sharing similar semantic properties tend to have high similarities~\cite{urban_anomaly}. For example, in Fig.~\ref{fig:challenges}, the adjacency of Region 1 and Region 2 suggests interconnected roads, while the similar distribution of POIs in Region 1 and Region 3 contributes to similar accident patterns. Effectively capturing both types of correlations poses a significant challenge.

\noindent (3) \textbf{Sparsity: Zero-Inflation Problem.} As shown in the lower right part of Fig.~\ref{fig:challenges}, the number of traffic accidents in NYC during an afternoon is presented in a heatmap. Notably, most regions experience zero traffic accidents. This is identified as the zero-inflation problem. Failure to properly address this sparsity can lead the model to generate predictions with an excessive number of zeros, rendering the predictions meaningless~\cite{Zero-inflated}. 

Unfortunately, existing studies have not thoroughly captured most of the above aspects. Traditionally, statistics-based methods~\cite{RWdectre,RWsvm,RWkneibor,RWnbr} merely utilize the statistics information of historical accidents, ignoring the influence of many spatio-temporal factors, which leads to poor performance. To alleviate this, learning-based methods~\cite{chen2018sdcae,hetero,DBLP:conf/aaai/WangL0W21,MVMT,C-ViT,hintnet, MG-TAR, RiskContra} have been proposed. They analyze spatio-temporal features to capture the correlations between traffic accidents. However, they fail to effectively capture the spatial proximity and semantic similarity of traffic accidents and miss out on the regional background of where they occur. Worse still, their approach to dealing with sparsity is not sufficiently effective.

To incorporate all the above critical aspects, we design a \underline{M}ulti-\underline{G}ranularity \underline{H}ierarchical \underline{S}patio-\underline{T}emporal \underline{N}etwork (\modelName) for traffic accident risk prediction. 

First, recognizing the \textbf{informative potential of remote sensing images in revealing regional backgrounds}, we integrate them into the traffic accident risk prediction process. This innovative addition serves a dual purpose: enhancing the spatio-temporal feature set and aiding in semantic analysis. We achieve this by converting remote sensing images into spatial features, augmenting the data. Simultaneously, we introduce a self-supervised autoencoder to encode these images for enhanced semantic understanding.

Second, \textcolor{black}{to capture geographical correlations, following the first law of geography in geography science, we \textbf{utilize spatio-temporal region features that encompass the geographical layout of the regions} to capture the physical proximity between different areas.} Additionally, \textcolor{black}{to capture traffic correlations,} we \textbf{construct a multi-view semantic similarity graph that quantifies the semantic similarities among regions} based on attributes like Points of Interest (POI) distributions. By integrating these two components, we lay the foundation for a model that inherently considers both spatial and semantic aspects. 

Third, to address the zero-inflation problem, motivated by the idea of aggregation, we \textbf{enhance the risk prediction at the lowest level by introducing multiple high-level risk prediction tasks}, carefully tailored to low-sparsity data settings. In particular, we use \textcolor{black}{regularized region partition} and graph clustering to respectively construct hierarchical structures based on spatio-temporal region features and semantic graph features. In addition, we design a multi-level embedding fusion mechanism to fully leverage the hierarchical structure for mutual learning of different levels. Furthermore, we devise an adaptive temporal attention module to acquire temporal dependencies from the representations encoded by preceding modules, effectively harnessing the attention mechanism to adeptly capture correlations. Finally, we design a multivariate loss function to learn the proposed model comprehensively.

In summary, we make the following contributions:
\vspace{-\topsep}
\begin{itemize}[leftmargin=10.2pt]
\setlength{\itemsep}{0pt}
\setlength{\parsep}{0pt}
\setlength{\parskip}{0pt}
\item We design a multi-granularity hierarchical spatio-temporal network, \modelName, that can fully exploit spatio-temporal features in traffic accidents in hierarchical manner. (Sec.~\ref{sec:3})
\item We incorporate remote sensing images to enhance regional background comprehension and create multi-level hierarchical structures for both region feature and multi-view graph. (Sec.~\ref{sec:3.1})
\item We design multiple encoding modules for multi-source spatio-temporal features and propose a cross-level message passing module to enable mutual learning across different levels. (Sec.~\ref{sec:3.2})
\item We comprehensively model the traffic accident risk prediction problem by designing a multivariate loss function to cater to the multi-faceted objective of prediction. (Sec.~\ref{sec:3.3})
\item We conduct a comprehensive evaluation on two real-world datasets. The results show that our method significantly outperforms state-of-the-art in terms of accuracy and robustness. (Sec.~\ref{sec:4})
\end{itemize}

\begin{figure}
  \centering
  \includegraphics[width=0.95\linewidth]{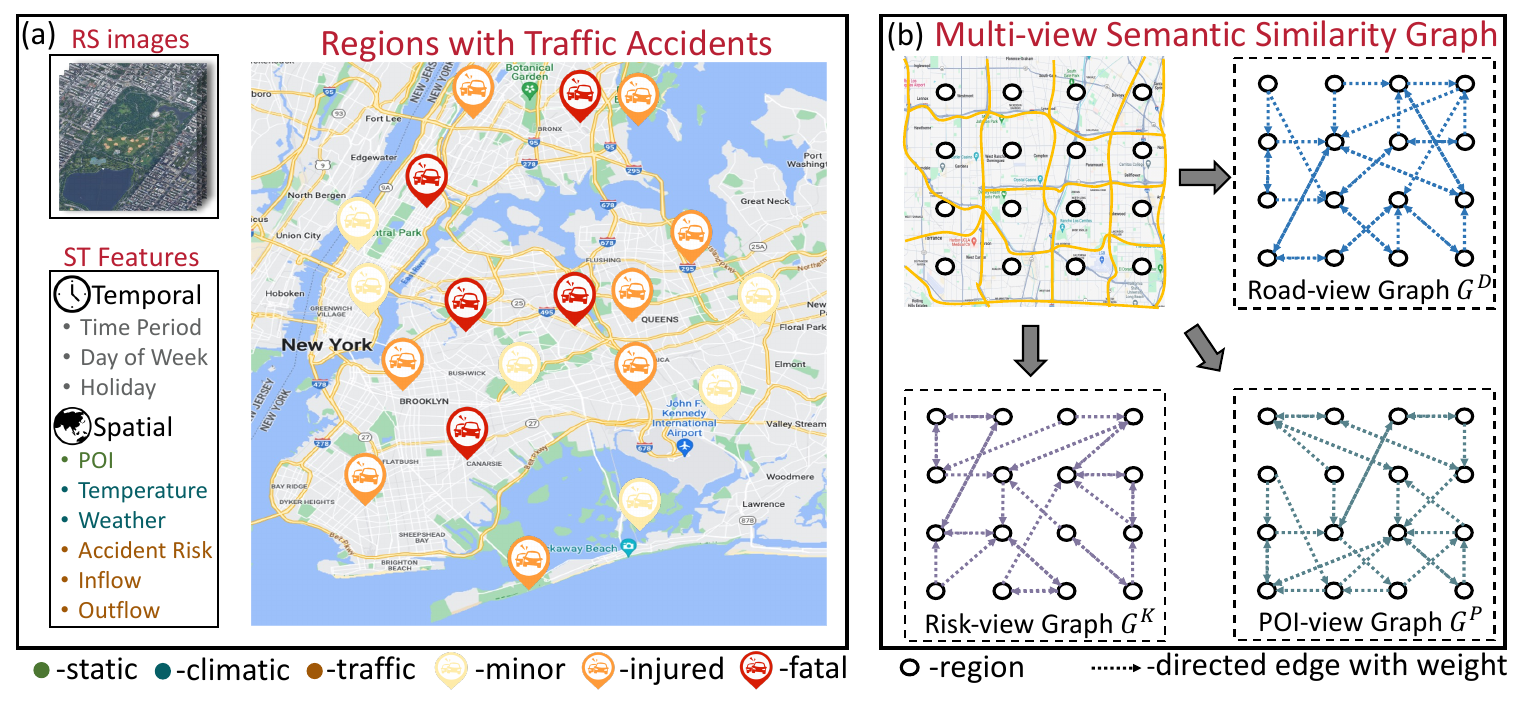}
  \vspace{-0.15in}
  \caption{An illustration of preliminaries, (a) shows the region features input and (b) shows the graph features input.}
  \vspace{-0.2in}
  \label{fig:pre}
\end{figure}

\section{PROBLEM FORMULATION}
In this section, we first introduce some useful preliminaries. Then we formalize the traffic accident risk prediction problem.

\subsection{Preliminaries}

\textcolor{black}{\noindent\textbf{Region.} In accordance with established urban zoning, the city is divided into $N$ regions based on administrative divisions. Each of these regions is identified with an index $i$, and is referred to as $U_i$. This division improves traffic accident risk prediction by enabling detailed analysis of each region.}

\noindent\textbf{Remote Sensing Images.}
Remote sensing has become an effective tool in understanding urban environments, which can provide rich information about topographic features of urban areas. As shown in Fig.~\ref{fig:pre}~(a), we introduce remote sensing images $\mathbf{X} = \{x_{i}\}_{i=1}^{N_{rs}}$ to enhance regional background comprehension, where $x_{i} \in \mathbb{R}^{W\times H \times d_x}$.

\noindent\textbf{Traffic Accident Risk Map.} 
As shown in Fig.~\ref{fig:pre}~(a), we categorize traffic accidents based on the severity level in the dataset and assign a corresponding risk value to each category. Specifically, the risks of minor, injured, and fatal accidents have values of 1, 2, and 3, respectively. The traffic accident risk $A_{i}^t$ for region $U_{i}$ at interval $t$ is the sum of risk values of all traffic accidents within that region and interval. 
Therefore, the risk map of all regions at $t$ is denoted as the risk map $A^t \in \mathbb{R}^N$.

\noindent\textbf{Spatio-temporal Region Features.} 
As shown in Fig.~\ref{fig:pre}~(a), we construct multi-source spatio-temporal region features. The temporal features consist of hour of day, day of week, and holiday information, which can be easily obtained in advance. Since they are the same for all regions, we extend their dimension to fit the form of regions, which can be denoted as $\mathcal{T} \in \mathbb{R}^{N \times d_t}$, where $d_{t}$ is the dimension of temporal features. The spatial features $\mathcal{S} \in \mathbb{R}^{N \times d_{s}}$ encompass static features, climatic features and traffic features, where $d_{s}$ is the dimension of spatial features. Specifically, static features contain the distribution of different types of POI, climatic features contain temperature and weather information, and traffic features contain traffic accident risk map, inflow and outflow of each region. Combining the spatial and temporal features, we use $ST \in \mathbb{R}^{N \times d_{st}}$ to represent spatio-temporal features, where $d_{st} = d_t + d_s$.

\noindent\textbf{Multi-view Semantic Similarity Graph.}
Intuitively, regions with similar semantic properties may have similar traffic accident patterns. To fully capture the semantic similarity between regions, as shown in Fig.~\ref{fig:pre}~(b), we construct a multi-view semantic similarity graph $G^t=(V,E,M^t)$ at each interval $t$. Specifically, we use traffic features (i.e., traffic accident risk map, inflow, outflow) during the interval $t$ as node features $M_n^t$, which means the graph is dynamic over time. Graph $G^t$ contains three views, the road-view, the risk-view and the POI-view. The road view highlights similarities in road patterns, including road type, lane structure, and traffic signs. The risk view focuses on similarities in risk features. The POI view shows similarity in POI distribution, characterized by type and quantity. Note that the three views share the same nodes $V$ and node features $M^t$, but different edges $E=\{E_D,E_K,E_P\}$. The construction detail is explained in Sec.~\ref{sec:3.1.3}.

\subsection{Problem Definition}
Given the Regions $\mathbf{U}$, their corresponding remote sensing images $\mathbf{X}$, historical spatio-temporal region features $\{ST^1, ST^{2}, \cdots, ST^T\}$, historical semantic similarity graphs $\{G^1, G^{2}, \cdots,G^{T}\}$ and the future temporal feature $\mathcal{T}^{T+1}$, our goal is to predict the traffic accident risk map $\hat{A}^{T+1}$ at the next interval $T+1$.



\begin{figure}
  \centering
  \includegraphics[width=\linewidth]{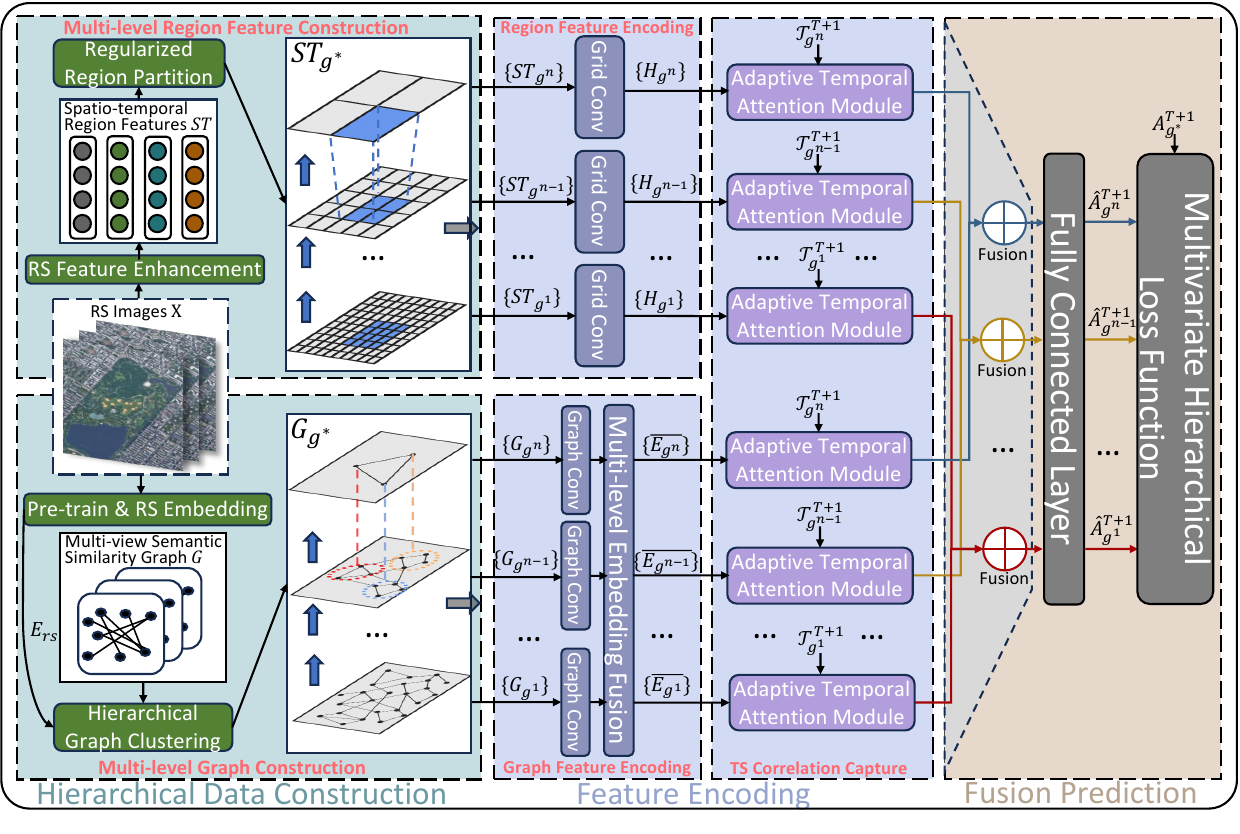}
  \vspace{-0.25in}
  \caption{The model framework of \modelName.}
  \vspace{-0.15in}
  \label{fig:model}
\end{figure}
\section{METHODOLOGY}
\label{sec:3}
In this section, we present our proposed \textbf{\modelName}. As shown in Fig.~\ref{fig:model}, it consists of three key phases: \textit{Hierarchical Data Construction} in Sec.~\ref{sec:3.1}, \textit{Feature Encoding} in Sec.~\ref{sec:3.2}, and \textit{Fusion Prediction} in Sec.~\ref{sec:3.3}. The structure constructed in the first phase enhances the model's ability to address sparsity issues from a novel hierarchical perspective. In the second phase, effective representation learning is meticulously carried out for the constructed hierarchy. Finally, the third phase fully integrates the feature representations and delivers comprehensive and effective prediction results. Each phase incorporates innovative strategies that significantly enhance the model's performance in estimating traffic accident risk.

\subsection{Hierarchical Data Construction}
\label{sec:3.1}
In this section, we innovatively constructs two types of hierarchical structures for region feature and graph feature. Moreover, by integrating remote sensing encoding methods tailored to each type of feature, we construct a more subtle hierarchical data structure with effective regional background comprehension.

\subsubsection{Remote Sensing Feature Enhancement}
\label{sec:3.1.1}

To enhance regional background understanding through remote sensing images, we first use the CNN module to encode the remote sensing images \(\mathbf{X}\), which can be formulated as the following equation:
\begin{equation}
\begin{small}
\begin{aligned}
    X_k &= MaxPool(\sigma(W_k * X_{k-1} + b_k))) \\
    F_{rs} &= FC(X_{k})
    \end{aligned}
\end{small}
\end{equation}
where $*$ represents convolution operation, $\sigma(\cdot)$ is the ReLU activation function, $k$ is the layer of convolution. $W_k , b_k$ are trainable weights. 
$F_{rs} \in \mathbb{R}^{N \times d_{a}}$ is the encoded remote sensing feature.

The spatio-temporal region features $ST$ we originally constructed already contain multivariate information, which is the foundation for our model to fully learn the complex spatio-temporal correlations between traffic accidents. 
To further improve regional background comprehension, we use the encoded remote sensing features for enhancement, which are concatenated with $ST$.

\subsubsection{Multi-level region feature construction}
\label{sec:3.1.2}

The original spatio-temporal region features $ST \in \mathbb{R}^{N \times d_{st}}$ are divided by regions, to capture localized geographical correlations, we use different sized grids to subdivide regions, resulting in multi-level hierarchical spatio-temporal region features $\{ST_{g^1}, ST_{g^2}, \ldots, ST_{g^n}\}$, where $g^1$ denotes the finest granularity, $g^n$ represents the coarsest granularity. Specifically, constrained by the granularity of data collection, we aggregate the finest granularity data to obtain coarser-grained hierarchical data.
Considering $ST$ contains multi-source features, which have different statistical characteristics, therefore we utilize the \textit{maximum}, \textit{mean} and \textit{summation} methods respectively for aggregation. For instance, the maximum method can ensure that no weather conditions are lost after aggregation, the average method can determine the appropriate temperature of aggregated grids, and the summation method can reasonably obtain the aggregated value for traffic accident risk.

\subsubsection{Multi-view Graph Construction}
\label{sec:3.1.3}
Traffic accidents show a connection that can be largely understood by analyzing semantic features on the ground, such as POIs and road structures~\cite{DBLP:conf/gis/LiangJZ17}. Based on the semantic features, we construct three semantic similarity graphs, namely the road-view similarity graph $G^D=(V,E^D)$, the risk-view similarity graph $G^K=(V,E^K)$ and the POI-view similarity graph $G^P=(V,E^P)$. 

To represent the similarity of road, risk, and POI between any two nodes, we use Jensen-Shannon divergence~\cite{DBLP:journals/tit/Lin91} to calculate the similarity score. Taking the POI similarity as an example, it can be computed by the following equation:
\begin{equation}
\begin{small}
\begin{aligned}
    Sim_{P}(U_{i}, U_{j}) = 1 - JS(R_P^{i}, R_P^{j}) 
\end{aligned}
\end{small}
\end{equation}
\begin{equation}
\begin{small}
\begin{aligned}
    JS(R_P^{i}, R_P^{j}) = \frac{1}{2}\sum_{d=1}^D(\log\frac{2R_P^{i}(d)R_P^{i}(d)+2R_P^{j}(d)R_P^{j}(d)}{R_P^{i}(d)+R_P^{j}(d)})
\end{aligned}
\end{small}
\end{equation}
where $R_P^{i}, R_P^{j}$ denote the POI distribution of region $U_{i}$ and $U_{j}$ and $D$ is the dimension of POI information. 
Similarly, we also calculate $Sim_{D}(U_{i}, U_{j})$ and $Sim_{K}(U_{i}, U_{j})$ for road-view and risk-view.

Afterwards, we use the semantic similarity calculated between each pair of nodes as the weight and select the Top-$K$ most similar nodes as neighbors for each node. Finally, we construct adjacency matrices $M_A=[M_A^{D}, M_A^{K}, M_A^{P}] \in \mathbb{R}^{N \times N}$ for three different views, where $N$ is the number of regions.

The adjacency matrices represent global semantic similarity structures between regions, but has no specific node features. Therefore, we use traffic features (i.e., traffic accident risk map, inflow, outflow) as node features $M_n^t \in \mathbb{R}^{N \times d_c}$ for each time interval $t$. Finally, we perform matrix multiplication between the adjacency matrices $M_A$ of the three views and the node features $M_n^t$ separately to get multi-view semantic similarity graph $G^t = M_A \cdot M_n^t$.


\begin{figure}
  \centering
  \includegraphics[width=0.9\linewidth]{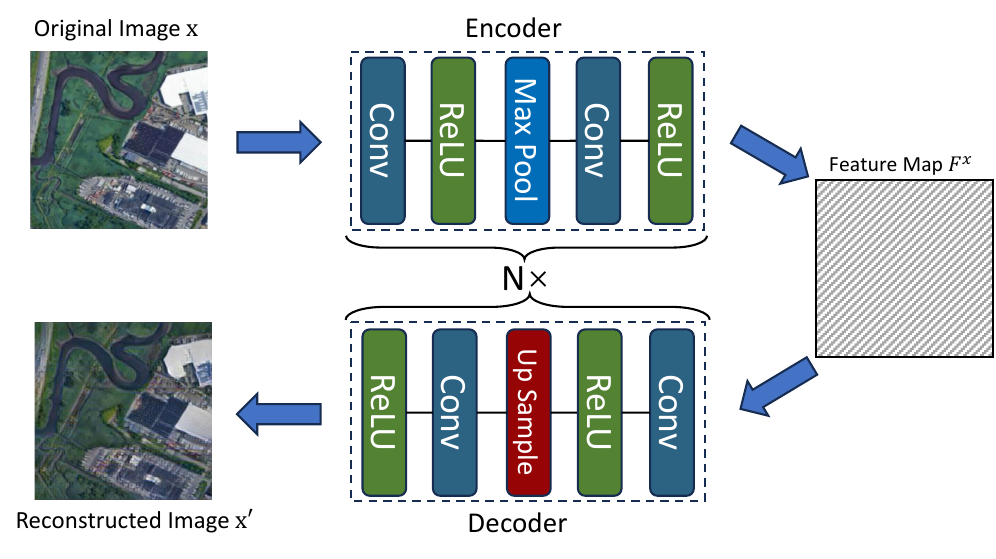}
  \vspace{-0.1in}
  \caption{The Framework of Remote Sensing Autoencoder}
  \vspace{-0.1in}
  \label{fig:AE}
\end{figure}

\subsubsection{Pre-train \& Remote Sensing Embedding}
\label{sec:3.1.4}
Remote sensing images can provide regional information, which can be encoded as remote sensing features $F_{rs}$ as done in Sec.~\ref{sec:3.1.1}. For multi-level graph construction, we further mine them from another perspective, the perspective of similarity clustering.
Aiming to learn the semantic properties of remote sensing images, autoencoder can be trained as a feature extractor by minimizing the reconstruction error between input and output data~\cite{kramer1991nonlinear}. To obtain remote sensing embeddings as much as possible from the global perspective, as shown in Fig.~\ref{fig:AE}, we propose a remote sensing autoencoder. The entire process of remote sensing autoencoder is as follows:
\begin{equation}
\begin{small}
\begin{aligned}
    F^x_{k} &=\sigma(MaxPool(\sigma(W_k^{(E_0)}*F^x_{k-1} + b_k^{(E_0)}))*W_k^{(E_1)}+b_k^{(E_1)})
  \\
    x'_{k} &=\sigma(UpSample(\sigma(W_k^{(D_0)}*x'_{k-1} + b_k^{(D_0)}))*W_k^{(D_1)}+b_k^{(D_1)})
  \end{aligned}  
    \end{small}
\end{equation}
where $*$ represents convolution operation, $F^x_{k}$ is the feature map of $k$-th encoder, $\sigma(\cdot)$ is the ReLU activation function, $W_k$, $b_k$ are trainable weights, 
$x'_k$ is the reconstructed image of $k$-th decoder.

In this work, we pre-train the autoencoder from scratch by using two different losses, which can be computed as follows:
\begin{equation}
\begin{small}
\begin{aligned}
    loss_{pp}(x,x') &=\frac{1}{c' \cdot h' \cdot w'}\cdot \sum_{i=1}^{c'}\sum_{j=1}^{h'}\sum_{k=1}^{w'}(x_{ijk}-x'_{ijk})^2 \\
    loss_{feat}(F_{ij}^{x},F_{ij}^{x'}) &=\frac{1}{c' \cdot h' \cdot w'}\cdot \sum_{i=1}^{c}\sum_{j=1}^{h\cdot w}(F_{ij}^{x}-F_{ij}^{x'})^2
\end{aligned}
\end{small}
\end{equation}
where $loss_{pp}$ is the per-pixel loss, forces the pixel value of $x'$ to resemble the ones of $x$. $F^{x}$ and $F^{x'}$ are the feature maps obtained by encoding input image $x$ and reconstructed output image $x'$. $loss_{feat}$ is the feature loss, measures the difference between the feature maps $F^{x}$ and the feature maps $F^{x'}$.
After pre-training, we separately use the encoder part to obtain remote sensing embeddings $E^{rs}$

\subsubsection{Multi-level Graph Construction.}
\label{sec:3.1.5}
To construct hierarchical graph structure, we propose Algorithm~\ref{alg:HGC} for hierarchical graph clustering to get hierarchical aggregation relationships $\mathcal{R}_a$.

\begin{algorithm}
\small{
    \KwIn{RS embedding $E^{rs}$, part number $[\mathcal{N}_{1}, \mathcal{N}_{2}, \cdots, \mathcal{N}_{n}]$, graph size $[\mathcal{L}_{1}, \mathcal{L}_{2}, \cdots, \mathcal{L}_{n}]$, granularity number $n$}
    \KwOut{hierarchical aggregation relationships $\mathcal{R}_a$}
    \caption{Hierarchical Graph Clustering}
    \label{alg:HGC}
    build \textit{RS similarity graph} $G_{g^1}^{rs}$ at granularity $g^1$ with $E^{rs}$\;
    
    \For{$i \gets 1 \cdots n$}{
        $\mathcal{M}_i \gets graphPartition(\mathcal{N}_{i}, \mathcal{L}_{i}, G_{g^i}^{rs})$\;
        
        $\mathcal{P}_{i} \gets extract(\mathcal{M}_i)$\;
        
        $X_{i+1}^{rs} \gets averageAggregate(\mathcal{P}_{i}, X_{i}^{rs})$\;
        
        build next level \textit{RS similarity graph} $G_{g^{i+1}}^{rs}$ with $X_{i+1}^{rs}$\;
    }
    $\mathcal{R}_a \gets [\mathcal{P}_{1},\cdots,\mathcal{P}_{n}]$\;
    return $\mathcal{R}_a$
    }
\end{algorithm} 

Firstly, we construct a novel remote sensing similarity graph denoted as $G^{rs}=(V,E^{rs})$. We represent regions as nodes and use the remote sensing embedding $E^{rs}$ as node features. We then establish edges $E^{rs}$ between adjacent nodes based on spatial relationships. The weight of each edge between adjacent nodes is determined by the cosine similarity $cos(E^{rs}_{i},E^{rs}_{j})$.

Next, to classify nodes with similar remote sensing features and close spatial distances into the same category, we choose the classic graph clustering algorithm Metis~\cite{metis}. Its primary goal is to create partitions that are roughly equal in size while minimizing the number of "edges cut" (i.e., edges that cross between different parts). For the remote sensing similarity graph $G_{g^i}^{rs}$, we employ graph partition to obtain membership sets $\mathcal{M}_i$ for all subgraphs. We then extract node parts $\mathcal{P}_{i}$ from $\mathcal{M}_i$ and calculate the next level remote sensing embedding $E_{i+1}^{rs}$ based on $\mathcal{P}_{i}$ and $E_{i}^{rs}$.

Then, we construct $G_{g^{i+1}}^{rs}$ based on $E_{i+1}^{rs}$, and repeat the above procedures until $i=n$, where $n$ is number of granularity levels. We use the pre-computed part number list $[\mathcal{N}_{1}, \mathcal{N}_{2}, \cdots, \mathcal{N}_{n}]$ and graph size list $[\mathcal{L}_{1}, \mathcal{L}_{2}, \cdots, \mathcal{L}_{n}]$ for each level's graph partition.

Finally, we get the hierarchical aggregation relationships $\mathcal{R}_a$ by combining node parts $[\mathcal{P}_{1},\cdots,\mathcal{P}_{n}]$ from each level.

To construct the hierarchical graph structure, we take the original multi-view graph $G$ as the finest-grained basis and perform hierarchical aggregation on it according to $\mathcal{R}_a$. Specifically, we construct data for each granularity in a hierarchical manner from fine to coarse, which can be outlined as the following process:

\noindent  \textcircled{1} Firstly, based on the hierarchical aggregation relationship $\mathcal{R}_a$, we aggregate the node set $V_{g^i}$ of the last granularity $g^i$ to obtain a new node set $V_{g^{i+1}}$ for the current granularity $g^{i+1}$.

\noindent \textcircled{2} Then, the current granularity's graph inherit the connectivity of the last granularity's graph (e.g., for the current granularity's nodes $V_{g^{i+1}}^A$ and $V_{g^{i+1}}^B$ aggregated from last granularity's nodes $[V_{g^{i}}^a,\cdots]$ and $[V_{g^{i}}^b,\cdots]$, there is an edge between $V_{g^{i+1}}^A$ and $V_{g^{i+1}}^B$ as long as there is an edge between any $V_{g^{i}}^a \in V_{g^{i+1}}^A$ and any $V_{g^{i}}^b \in V_{g^{i+1}}^B$). 

\noindent \textcircled{3} Next, for edges at the current granularity, we use the maximum edge weight associated with it from the last level as its weight.

\noindent \textcircled{4} Finally, to maintain consistency with the initial graph, we only keep the top $K$ edges with highest weights for each node.

This process leads to the creation of semantically rich and hierarchically structured graphs.

\subsection{Feature Encoding}
\label{sec:3.2}
In this section, we address the distinctive characteristics of geography, traffic and time series by introducing dedicated modules for their encoding: the \textit{Region Feature Encoding} in Sec.~\ref{sec:3.2.1} to capture spatial proximity, the \textit{Graph Feature Encoding} in Sec.~\ref{sec:3.2.2} to capture semantic similarity and the \textit{Time Series Correlation Capture} in Sec.~\ref{sec:3.2.3} to capture time series dependencies.

\subsubsection{Region Feature Encoding}
\label{sec:3.2.1}
According to previous studies~\cite{DBLP:conf/aaai/ZhouWXCL20,DBLP:conf/aaai/WangL0W21,C-ViT,MVMT}, for a specific region, the traffic accident risk of its target time interval is highly correlated with several previous time intervals on the same day and the same time interval several weeks prior, which is known as the short-term proximity and long-term periodicity. To this end, for all granularity $g^*$, we fetch region features from the previous $p$ time intervals and the same time interval in previous $q$ weeks as the historical observation sequence, whose length is $T=p+q$. Considering the relationships between region features and traffic accidents are highly dependent on the geographical spatial correlation between grids, we utilize convolutions to capture this correlation, which can be formulated as:
\begin{small}
\begin{equation}
    H_{g}^{t,k}=\sigma(W_{g}^{k}*H_{g}^{t,k-1}+b_g^{k})
\end{equation}
\end{small}
where $*$ represents convolution operation, $\sigma(\cdot)$ is the ReLU activation function, $W_g^k$, $b_g^k$ are trainable weights, $H_{g}^{t,k}$ is the output of the $k$-th convolutional layer at time interval $t$ for granularity $g$. 
\subsubsection{Graph Feature Encoding}
\label{sec:3.2.2}

Firstly, similar to region feature, we construct a graph historical observation sequence consisting of short-term and long-term for each granularity. Afterwards, aiming to model the semantic correlation between regions, we leverage graph convolution to learn node representations for the semantic similarity graph $G_{g^*}=[G_{g^*}^D,G_{g^*}^K,G_{g^*}^P]$ at each granularity $g$, which is computed as follows:
\begin{small}
\begin{equation}
    E_{g}^{t}=\sigma(\sigma(G_{g}^{t}W_{g}^{(0)}+b_{g}^{(0)})W_{g}^{(1)}+b_{g}^{(1)}) 
\end{equation}
\end{small}
where $W_{g}$, $b_{g}$ are trainable weights, $\sigma(\cdot)$ is the ReLU activation function, $E_{g}^{t}$ is the embedding of semantic similarity graph. 

\begin{figure}
  \centering
  \includegraphics[width=0.9\linewidth]{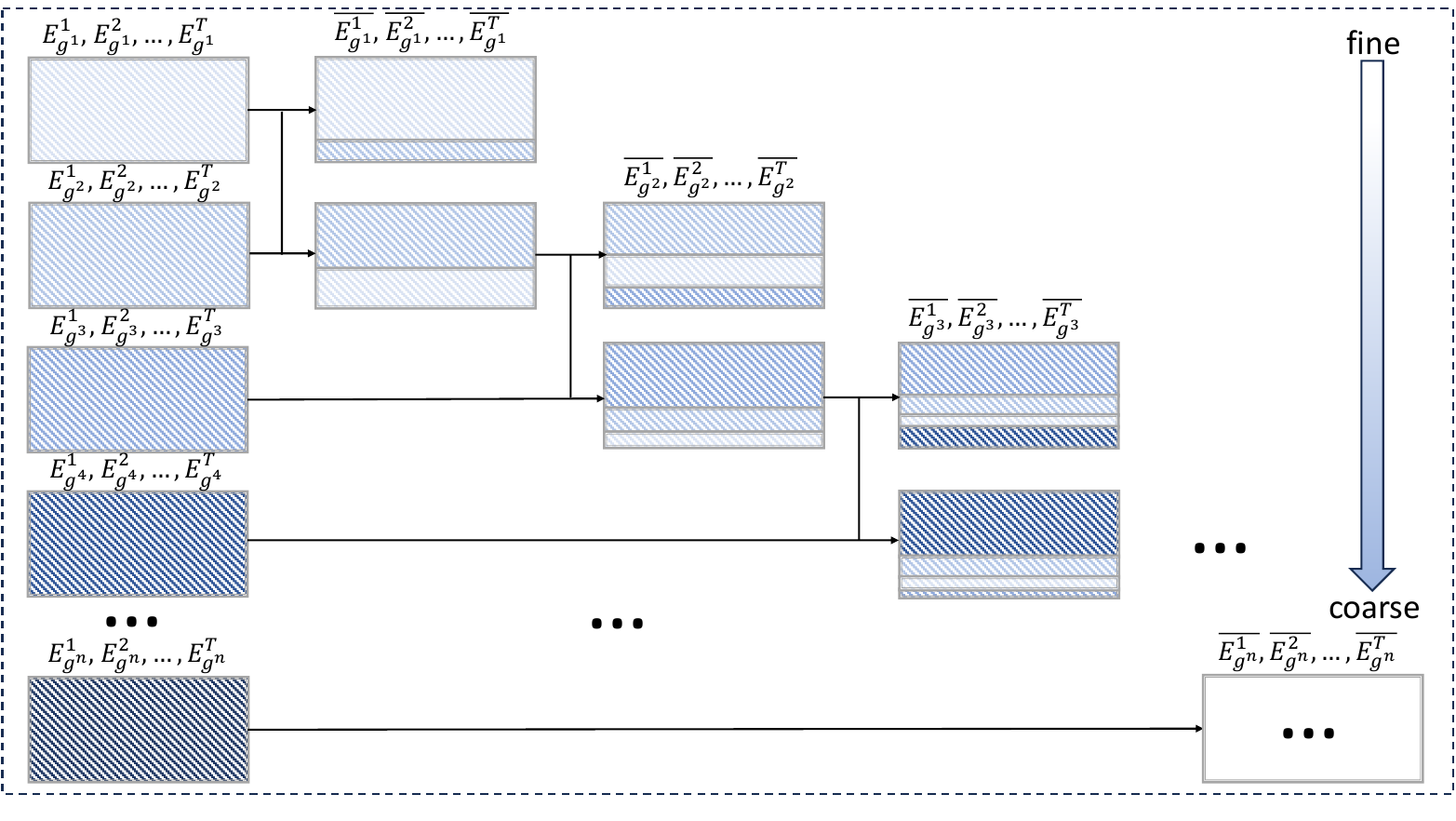}
  \vspace{-0.1in}
  \caption{Multi-level Embedding Fusion Mechanism}\label{fig:EF}
  \vspace{-0.1in}
\end{figure}

In Sec.~\ref{sec:3.1.5}, we perform hierarchical graph clustering based on remote sensing similarity, aiming to construct multi-level graph, which results in coarse-grained graph nodes having more accident counts. To enhance prediction at the finest-grained level and capture spatial relationships across varying granularities, we propose multi-level embedding fusion module. As shown in Fig.~\ref{fig:EF}, the information transfer between different granularity levels is based on the hierarchical aggregation relationship $\mathcal{R}_a$. Since each coarse-grained graph node is composed of several fine-grained graph nodes, there is an aggregation relationship between adjacent levels. Based on that, we propose the following procedure:
\begin{itemize}[leftmargin=15pt]
    \item [1.] Firstly, we construct a granularity transformation matrix $M_{tran}$ for any two adjacent layers based on $\mathcal{R}_a$ as follows:
    \begin{small}
    \begin{equation}
        M_{tran}(i,j)=\left\{
                  \begin{array}{ll}
                    1 & \text{if}\: U^i_{g^{fine}} \in U^j_{g^{coarse}}\\
                    0 & \text{otherwise} 
                  \end{array}
                \right.
        \label{eq:Mtran}
    \end{equation}
    \end{small}
    where $U^i_{g^{fine}}$ denotes the region $i$ of granularity $g^{fine}$ and $U^j_{g^{coarse}}$ denotes the region $j$ of granularity $g^{coarse}$.
    \item [2.] Then, for any two adjacent layers, we use the following equation to perform embedding fusion:
    \begin{small}
    \begin{equation}
    \begin{aligned}
        \overline{E_{g^{coarse}}} & =E_{g^{coarse}}+\lambda_f M_{tran}^TE_{g^{fine}}
        \\ 
        \overline{E_{g^{fine}}} & =E_{g^{fine}}+\lambda_c M_{tran}E_{g^{coarse}}
        \end{aligned}
    \end{equation}
    \end{small}
    where $\lambda_f$ and $\lambda_c$ are the fusion coefficient of relatively fine layer and coarse layer.
    \item [3.] Finally, we repeat step 2 from fine to coarse, generating fused embedding $\overline{E_{g}}$ for each granularity level. The output sequence of this module is denoted as $\{\overline{E}\} = \{\overline{E^1},\overline{E^2},\cdots,\overline{E^T}\}$
\end{itemize}

\begin{figure}
  \centering
  \includegraphics[width=0.9\linewidth]{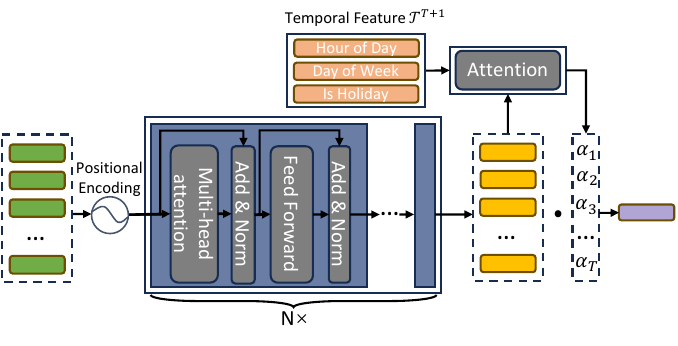}
  \vspace{-0.15in}
  \caption{Illustration of Adaptive Temporal Attention Module}\label{fig:TA}
  \vspace{-0.15in}
\end{figure}
\subsubsection{Time Series Correlation Capture}
\label{sec:3.2.3}
So far, the model has completed preliminary feature encoding. To fully capture the short-term proximity and long-term periodicity, as shown in Fig.~\ref{fig:TA}, we propose an adaptive temporal attention module. Specifically, taking the output sequence $\{H\}$ of the \textit{region feature encoding} as an example, we first follow the mechanism proposed in~\cite{DBLP:conf/nips/VaswaniSPUJGKP17} to perform positional encoding. For each element $H^t$ in the sequence, it corresponds to a positional encoding $H_o^t \in \mathbb{R}^{d_{st}}$, which is computed as follows:
\begin{equation}
\begin{small}
\begin{aligned}
    H_o^t[2k] &=sin(t/10000^{\frac{2k}{d_{st}}}) \\
    H_o^t[2k+1] &=cos(t/10000^{\frac{2k}{d_{st}}})
    \end{aligned}
    \end{small}
\end{equation}
where $H_o^t[2k]$ and $H_o^t[2k+1]$ correspond to even and odd dimensions of $H_o^t$, respectively. 
Next, we apply $N$ self-attention blocks to convert the code $H^t$ into the code $\overline{H}^t$ for each time interval $t$. Specifically, each block includes the following steps:
\begin{itemize}[leftmargin=15pt]
    \item [1.]\textbf{Self-attention:} For each code $H^t$, we first generate $query(Q^t = W_qH^t)$, $key(K^t = W_kH^t)$ and $value(V^t = W_vH^t)$. Then, we compute the $score(t,t') = softmax(\frac{Q^tK^{t'}}{\sqrt{d_{st}}})$. Afterwards, the code $H^t$ would be converted into a new representation (i.e., $z_a^t=\sum_{t'}score(t,t')V^{t'}$).
    
    \item [2.]\textbf{Add \& Normalize:} We leverage ResNet~\cite{DBLP:conf/eccv/HeZRS16} and the layer-normalization~\cite{DBLP:journals/corr/BaKH16} to accelerate the training process. Hence, the code would be updated: $\overline{H}^t = LN(\overline{H}^t \oplus z_a^t)$

    \item [3.]\textbf{Feed Forward:} For each updated code, we further encode it with a fully connected layer: $z_f^t = ReLU(W_f\overline{H}^t+b_f)$. 

    \item [4.]\textbf{Add \& Normalize:} Similar to step 2, the code is updated as following: $\overline{H}^t = LN(\overline{H}^t \oplus z_f^t)$
\end{itemize}
For simplicity, we denote the above four steps in the $i$-th block for granularity $g$ as $SAB_g^i(\cdot)$. Thus, the sequence $\{\overline{H}_g\}$ is generated by $N$ self-attention blocks, which can be formulated as below:
\begin{small}
\begin{equation}
    \{\overline{H}_g\} = SAB_N(SAB_{N-1}(\cdots SAB_1(\{H_g\})))
\end{equation}
\end{small}

Considering that the sequence $\{H\}$ is composed of historical observations from different time intervals, each element $H^t$ in sequence has different impact on the traffic accident situation at target time interval $T+1$, we introduce a temporal attention mechanism to adaptively capture the dynamic correlation between historical observations and target interval. Specifically, we first calculate the attention scores between $\{\overline{H}\}$ and target time interval's temporal feature $\mathcal{T}^{T+1}$, and use them as weights for summation to obtain the final output $\{\hat{H}\}$, which can be formulated as follows:
\begin{equation}
\begin{small}
\begin{aligned}
    \alpha = softmax(ReLU(\{\overline{H}\}W_H + \mathcal{T}^{T+1}W_T+b_{\alpha}))
\end{aligned}
\end{small}
\end{equation}
\begin{equation}
\begin{small}
\begin{aligned}
\{\hat{H}\}=\sum_{i=1}^T\alpha_i\cdot\overline{h}^i
\end{aligned}
\end{small}
\end{equation}
where $W_H$, $W_T$ and $b_{\alpha}$ are trainable weights, $\alpha \in \mathbb{R}^T$ is the attention score vector, which can be considered as the importance distribution of different historical observation. Similarly, for the output sequence $\{\overline{E}\}$ of the cross-granularity message passing part, the same adaptive temporal attention module is adopted to obtain $\{\hat{E}\}$.

\subsection{Fusion Prediction}
\label{sec:3.3}
First, we fuse each pair of $\{\hat{H}_{g}\}$ and $\{\hat{E}_{g}\}$, which are the final encoding of region features and graph features. Considering that they are obtained from different form of data, thus have different degrees of influence on the target region, we therefore adapt two trainable weight matrices and a fully connected layer to dynamically fuse them, which can be formulated as below:
\begin{equation}
\begin{small}
    \hat{A}_g^{T+1} = FC(W_1\{\hat{H}_{g}\}+W_2\{\hat{E}_{g}\})
    \end{small}
\end{equation}
where $W_1$ and $W_2$ are trainable weights, and $\hat{A}_g^{T+1}$ is the traffic accident risk map prediction of granularity $g$.

Next, to model the problem comprehensively, we design a multivariate hierarchical loss function, which consists of three parts: weighted mean squared error $loss_{wmse}$, binary cross entropy $loss_{bce}$, and hierarchical constraint $loss_{hc}$:

\noindent\textbf{Weighted Mean Squared Error:} Motivated by~\cite{DBLP:conf/aaai/WangL0W21}, to give higher emphasis to high-risk accident areas and address the issue of zero inflation, we classify all accident samples into four levels based on their risks and assign different weights to them:
\begin{small}
\begin{equation}
    loss_{wmse} = \frac{1}{N_g}\sum_{i \in I}\lambda_i(A_g(i)-\hat{A}_g(i))^2
\end{equation}
\end{small}
where $\hat{A}_g$ and $A_g$ are the prediction and ground truth of granularity $g$, $A_g(i)$ and $\lambda_i$ are the samples and weight whose traffic accident risk level is $i$, and $N_g$ is the number of regions of granularity $g$.

\noindent\textbf{Binary Cross Entropy:} Different from the global perspective in $loss_{wmse}$, we utilize binary cross entropy to measure the accuracy of predicting the occurrence of traffic accidents:
\begin{small}
\begin{equation}
    loss_{bce} = -\frac{1}{N_g}\sum_{i=1}^{N}(A_g^{i}\log \hat{A}_g^{i} + (1-A_g^{i})\log (1-\hat{A}_g^{i}))
\end{equation}
\end{small}
where $\hat{A}_g^{i}$ and $A_g^{i}$ are the prediction and ground truth of region $U_{i}$.

\noindent\textbf{Hierarchical Constraint:} To fully utilizing our multi-granularity hierarchical structure, we construct hierarchical constraint by utilizing the consistency between adjacent granularity level prediction results. Specifically, only the constraint between the finest-grained level $g^1$ and second-fine-grained level $g^2$ are considered since their number of regions is sufficient, which can be formulated as follows:
\begin{small}
\begin{equation}
     loss_{hc} = \frac{1}{N_{g^2}}(A_{g^2}-\hat{A}_{g^1}M_{tran})
\end{equation}
\end{small}
where $A_{g^2}$ is ground truth of granularity $g^2$, $\hat{A}_{g^1}$ is the prediction of granularity $g^1$, $M_{tran}$ is the  granularity transformation matrix computed by equation \ref{eq:Mtran}.

By combining the three mentioned loss functions, the final loss function is defined as follows:
\begin{equation}
\begin{small}
    loss_{F} = \sum_{i=1}^{n}(\lambda_w^i loss_{wmse}^i + \lambda_b^i loss_{bce}^i) + \lambda_{hc} loss_{hc}
    \end{small}
\end{equation}
where $\lambda_w^i$ and $\lambda_b^i$ are the weights of loss at granularity $g^i$, $\lambda_{hc}$ and $loss_{hc}$ are the weight and hierarchical constraint.

\section{EXPERIMENTS}
\label{sec:4}
\subsection{Datasets}
As shown in Table~\ref{tab:Datasets}, we use two large public real-world datasets collected from NYC\footnote{https://opendata.cityofnewyork.us/} and Chicago\footnote{https://data.cityofchicago.org/}. 
The traffic accident data contains date, time, latitude and longitude, and the number of causalities. The taxi order data includes location and time of pick-up and drop-offs, which is used to calculate the traffic flow. The POI data has seven categories: residence, school, culture facility, recreation, social service, transportation and commercial. The weather data contains temperature and sky condition (i.e., sunny, rainy, cloudy, snowy and foggy). Note that for Chicago, POI data is lacking, so we only construct the road-view and risk-view semantic similarity graph. The road segment data includes road types, length and width, and snow removal priority. For remote sensing image data, we get the images of corresponding time from Google Earth. 
\begin{table}[!t]
\centering
\caption{Statistics of Datasets}
\vspace{-0.15in}
\label{tab:Datasets}
\scalebox{0.8}{%
\begin{tabular}{|c|c|c|}
\hline
Dataset & NYC & Chicago\\
\hline
Time Span & 1/1/2013-12/31/2013 & 2/1/2016-9/30/2016\\
Accidents & 147K & 44K\\
Taxi Orders & 173,179K & 1744K\\
POIs & 15625 & None\\
Hours of Weather & 8760 & 5832\\
Road Segments & 103K & 56K\\
RS Images & 400 & 400\\
\hline
\end{tabular}
}
\vspace{-0.1in}
\end{table}

\subsection{Evaluation Metrics}
To evaluate the performance of our model, we utilise three commonly used metrics for traffic accident risk prediction task~\cite{DBLP:conf/cikm/MaZWL18,DBLP:conf/aaai/WangL0W21,C-ViT,MVMT,RiskContra}, namely RMSE, Recall and MAP. From the perspective of regression, RMSE is used to evaluate the overall prediction of traffic accident risk. From the perspective of ranking and classification, we use Recall to evaluate classification accuracy for accident area prediction, and use MAP to evaluate ranking accuracy for high accident risk areas prediction. Lower RMSE indicates that the model can predict risk more accurately overall, while a higher Recall and MAP indicate the better performance in high-risk regions, which can be considered having better coping abilities for zero-inflation problem. The three metrics are computed as follows: $RMSE = \sqrt{\frac{1}{T}\sum_{t=1}^T(A_t-\hat{A}_t)^2}$, $ Recall = \frac{1}{T}\sum_{t=1}^T\frac{S_t \cap R_t}{|R_t|}$, $ MAP = \frac{1}{T}\sum_{t=1}^T\frac{\sum_{j=1}^{|R_t|}pre(j)\times rel(j)}{|R_t|}$, where $A_t$ is the ground truth and $\hat{A}_t$ is the predicted values at time interval $t$. $R_t$ is the set of regions where traffic accidents have actually occurred at time interval $t$. $S_t$ is a set of regions with top $|R_t|$ highest predicted risks. $pre(j)$ denotes the precision of a cut-off rank list from 1 to $j$. $rel(j)=1$ if there are traffic accidents in region $j$, otherwise $rel(j)=0$.

Furthermore, we record the above metrics specifically on the rush hours during which the frequency of traffic accidents are usually higher, i.e., 7:00-9:00 and 16:00-19:00, and for simplicity we name them as RMSE*, Recall*, and MAP*, respectively. 

\subsection{Experimental Settings}
We partition all data into training, validation and test set with the ratio of 6:2:2. The city region is divided into grids with a size of $1982.5m\times 2776.5m$ for NYC and $2315m\times 2828m$ for Chicago. The time interval is one hour. The length of short-term $p$ and long-term $q$ are set to 3 and 4.
The number of granularity levels is set to 4.
In the weighted mean squared error, the weights $\lambda_i$ are respectively set to 0.05, 0.2, 0.25 and 0.5. For the multi-level embedding fusion mechanism, $\lambda_f$ and $\lambda_c$ are set to 0.8 and 0.2, respectively.


All deep learning methods were implemented with PyTorch 1.12 and Python 3.8, and trained with a Tesla V100 GPU. The platform ran on Ubuntu 18.04. In addition, we used Adam~\cite{DBLP:journals/corr/KingmaB14} as the optimization method with the mini-batch size of 32. The learning rate is set as 0.0001, and the training epoch is set as 70. 
The code and data is available at \url{https://github.com/faceless0124/MGHSTN}. 

\subsection{Baseline Methods}
We compare our model with the following baselines.
\begin{itemize}[leftmargin=10pt]
    \item \textbf{Avg}: The traffic accident risk is predicted by averaging short-term and long-term historical observations.
    \item \textbf{GRU}~\cite{GRU}: Gated Recurrent Unit. It can model historical sequential data to capture temporal dependencies.
    \item \textbf{H-ConvLSTM}~\cite{hetero}: It combines CNN with LSTM and uses a sliding window to capture the heterogeneity of different regions.
    \item \textbf{SDCAE}~\cite{chen2018sdcae}: It captures the spatial features between different regions by stacking multiple denoise convolution layers.
    \item \textbf{GSNet}~\cite{DBLP:conf/aaai/WangL0W21}: A model that captures spatio-temporal correlations from both geographic and semantic aspects. 
    \item \textbf{MVMT-STN}~\cite{MVMT}: A recent multi-task learning framework which can predict fine- and coarse-grained citywide traffic accident risks simultaneously to alleviate the sparsity issue.
    \item \textbf{C-ViT}~\cite{C-ViT}: A recent model that re-formulates the traffic accident risk prediction problem as image regression problem, and uses a contextual vision transformer to model the task. 
    \item \textbf{HintNet}~\cite{hintnet}: A recent hierarchical knowledge transfer network uses spatial partitioning to separate sub-regions and leverages spatio-temporal and graph convolutions.
    \item \textbf{RiskContra}~\cite{RiskContra}: The state-of-the-art method that employs a contrastive learning framework with multi-kernel networks to enhance risk sample representation.
\end{itemize}

\begin{table*}[ht]
\centering
\caption{Performance comparison of different models.}
\vspace{-0.1in}
\label{tab:performance_all}
\scalebox{0.85}{%
\begin{tabular}{|c|c|c|c|c|c|c|c|c|c|c|c|c|}
  \hline
  \multirow{2}{*}{Model} & \multicolumn{6}{c|}{NYC} & \multicolumn{6}{c|}{Chicago} \\
  \cline{2-13}
  {} & RMSE & Recall & MAP & RMSE* & Recall* & MAP* & RMSE & Recall & MAP & RMSE* & Recall* & MAP* \\
  \hline
  Avg & 10.3243 & 29.42\% & 0.1336 & 9.4994 & 30.90\% & 0.1346 & 12.9581 & 16.58\% & 0.0622 & 10.2564 & 19.89\% & 0.0844 \\
  GRU & 8.6782 & 28.76\% & 0.1234 & 7.1463 & 30.97\% & 0.1321 & 11.5774 & 17.41\% & 0.0652 & 8.7592 & 19.63\% & 0.0701 \\
  H-ConvLSTM & 7.9126 & 30.22\% & 0.1454 & 7.2219 & 31.55\% & 0.1509 & 11.3002 & 17.82\% & 0.0698 & 8.8920 & 18.97\% & 0.0770 \\
  SDCAE & 7.9965 & 30.87\% & 0.1567 & 7.1654 & 31.05\% & 0.1546 & 11.4678 & 18.23\% & 0.0709 & 8.6412 & 20.17\% & 0.0924 \\
  GSNet & 7.6153 & 33.16\% & 0.1787 & 6.7760 & 34.15\% & 0.1769 & 10.9659 & 19.92\% & 0.0671 & 8.3305 & 21.12\% & 0.0745 \\
  MVMT-STN & 10.3335 & 33.75\% & 0.1884 & 9.5021 & 35.09\% & 0.1864 & 13.0719 & 20.75\% & 0.0828 & 10.2665 & 22.22\% & 0.0931\\
  C-ViT & 8.5015 & 30.59\% & 0.1632 & 7.0144 & 31.66\% & 0.1577 & 11.7730 & 19.94\% & 0.0787 & 8.2250 & 19.91\% & 0.0855\\
  HintNet & 8.7562 & 32.81\% & 0.1732 & 7.0819 & 34.21\% & 0.1687 & 11.5860 & 19.13\% & 0.0817 & 8.6250 & 22.01\% & 0.0895\\
  RiskContra & 7.4014 & 34.21\% & 0.1854 & 6.7207 & 35.31\% & 0.1914 & 10.3564 & 21.20\% & 0.0919 & 7.6165 & 22.17\% & 0.0817\\
  \textbf{\modelName} & \underline{6.8860} & \underline{34.59\%} & \underline{0.1985} & \underline{6.4511} & \underline{35.67\%} & \underline{0.1934} & \underline{7.4771} & 21.32\% & 0.0886 & \underline{5.9452} & 22.36\% & 0.0946\\
  \hline
  Avg+RS & 10.3243 & 29.42\% & 0.1336 & 9.4994 & 30.90\% & 0.1346 & 12.9581 & 16.58\% & 0.0622 & 10.2564 & 19.89\% & 0.0844 \\
  GRU+RS & 8.5912 & 29.81\% & 0.1284 & 7.2413 & 31.46\% & 0.1387 & 11.2961 & 18.21\% & 0.0702 & 8.3781 & 19.37\% & 0.0794 \\
  H-ConvLSTM+RS & 7.8906 & 30.39\% & 0.1424 & 7.0288 & 32.35\% & 0.1561 & 11.2702 & 18.07\% & 0.0723 & 8.6761 & 19.26\% & 0.0869 \\
  SDCAE+RS & 7.7135 & 31.13\% & 0.1603 & 7.0890 & 31.75\% & 0.1599 & 11.1701 & 18.70\% & 0.0769 & 8.6080 & 20.37\% & 0.1031 \\
  GSNet+RS & 7.5048 & 33.60\% & 0.1837 & 6.7290 & 35.37\% & 0.1835 & 10.8481 & 21.05\% & 0.0826 & 8.1437 & 22.63\% & 0.1065 \\
  MVMT-STN+RS & 10.2213 & 33.98\% & 0.1904 & 9.3032 & 35.24\% & 0.1813 & 13.0056 & \underline{21.62\%} & 0.0940 & 10.2112 & \underline{23.44\%} & \underline{0.1201} \\
  C-ViT+RS & 8.1772 & 29.60\% & 0.1612 & 7.2183 & 31.95\% & 0.1522 & 11.5853 & 20.37\% & 0.0862 & 8.5841 & 20.42\% & 0.0856\\
  HintNet+RS & 8.0123 & 32.69\% & 0.1802 & 6.6801 & 34.57\% & 0.1729 & 11.0810 & 20.83\% & 0.0839 & 8.3109 & 22.71\% & 0.0958\\
  RiskContra+RS & 7.2017 & 34.51\% & 0.1897 & 6.5314 & 35.52\% & 0.1921 & 10.0721 & 21.37\% & \underline{0.0992} & 7.4985 & 22.87\% & 0.1024\\
  \textbf{\modelName+RS} & \textbf{6.5011} & \textbf{34.74\%} & \textbf{0.1961} & \textbf{5.9947} & \textbf{35.86\%} & \textbf{0.1950} & \textbf{6.9211} & \textbf{21.71\%} & \textbf{0.1054} & \textbf{5.1098} & \textbf{23.46\%} & \textbf{0.1329}\\
  \hline
\end{tabular}
}
\end{table*}

\begin{figure}
  \centering
  \includegraphics[width=0.85\linewidth]{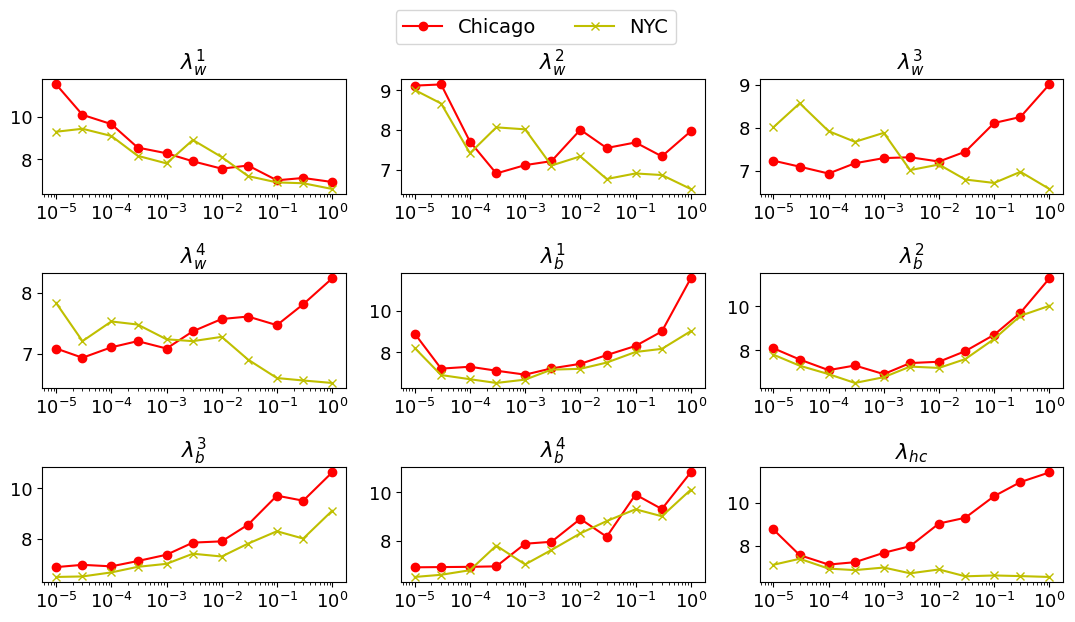}
  \vspace{-0.1in}
  \caption{RMSE for Chicago and NYC vs. the Loss Weight}\label{fig:weight}
  \vspace{-0.15in}
\end{figure}

\subsection{Effectiveness of Loss Weight}
To fine-tune the loss weight $\lambda_w^i$, $\lambda_b^i$ and $\lambda_{hc}$, we vary it from 0 to 1. We compute RMSE on the validation set, and the result is plotted in Fig.~\ref{fig:weight}. One can see that the loss weights have different impact on two datasets, and these three types of losses are effective for performance improvement, especially the WMSE of the finest-grained level and BCE.  Based on the results, (1) for Chicago, we set $\lambda_w^1 = 1,\lambda_w^2 = 3e-4,\lambda_w^3 = 1e-4,\lambda_w^4 = 3e-5,\lambda_b^1=1e-3,\lambda_b^2=1e-3,\lambda_b^3=1e-5,\lambda_b^4=1e-5,\lambda_{hc}=3e-4$, (2) for NYC, we set $\lambda_w^1 = 1,\lambda_w^2 = 1,\lambda_w^3 = 1,\lambda_w^4 = 1,\lambda_b^1=3e-4,\lambda_b^2=3e-4,\lambda_b^3=1e-5,\lambda_b^4=1e-5,\lambda_{hc}=1$.

\subsection{Effectiveness Comparison}
\label{sec:EC}
Since we innovatively introduce remote sensing data, while other methods do not, for fairness, we also introduce remote sensing data in the same way (i.e remote sensing enhancement) for the baseline methods. 
Table~\ref{tab:performance_all} shows the results of different models with and without remote sensing. we have the following observations:

\noindent(1) \textbf{Avg} is worse than any other deep learning based methods, since it does not utilize spatio-temporal features and cannot learn the spatio-temporal correlations of traffic accidents.

\noindent(2) \textbf{GRU} and \textbf{H-ConvLSTM} have relatively poor performance among learning-based methods. Although they outperform \textbf{Avg} on RMSE, they underperform \textbf{Avg} in some cases on Recall and MAP, indicating that they suffer from zero-inflation problem.


\noindent(3) Whether using remote sensing, \textbf{\modelName} outperforms other models overall. For example, \textbf{\modelName+RS} outperforms the best existing methods \textbf{RiskContra+RS} by 59.6\% on RMSE on Chicago, and also achieves significant improvement on Recall and MAP. Our model can make more accurate overall predictions while effectively dealing with the zero-inflation problem. 

\noindent(4) Remote sensing enhancement has improved the performance of nearly all methods, proving the effectiveness of introducing the regional background. The improvement is particularly obvious for Chicago. This may be because the addition of remote sensing images also alleviates the impact of the lack of POI data.

\noindent(5) The performance of all methods on NYC is better than that on Chicago. The reason is two-fold: first, Chicago lacks POI data compared to NYC; second, NYC dataset contains a higher number of accidents, which allows the model to learn better.

\begin{table}
\centering
\caption{Efficiency of Different Methods}
\vspace{-0.1in}
\label{tab:efficiency}
\scalebox{0.75}{%
\begin{tabular}{|c|c|c|c|}
  \hline
  \multirow{2}{*}{Model} & memory size & training time & prediction time\\
  {} & (NYC/Chicago) & (NYC/Chicago) & (NYC/Chicago)\\
  \hline
  Avg & 1.2GB/1.1GB & None & 0.01s/0.01s\\
  GRU & 8.4GB/10.3GB & 28.92s/20.38s & 0.05s/0.06s\\
  H-ConvLSTM & 9.5GB/10.9GB & 37.72s/28.44s & 0.07s/0.09s\\
  SDCAE & 9.1GB/10.2GB & 34.48s/27.47s & 0.06s/0.09s\\
  GSNet & 9.9GB/11.1GB & 37.32s/28.34s & 0.07s/0.09s\\
  MVMT-STN & 11.4GB/12.5GB & 48.79s/32.76s & 0.10s/0.11s\\
  C-ViT & 5.4GB/5.4GB & 15.84s/10.68s & 0.02s/0.02s\\
  HintNet & 10.3GB/11.7GB & 38.83s/30.47s & 0.07s/0.10s \\
  RiskContra & 10.8GB/11.9GB & 39.31s/32.59s & 0.09s/0.13s \\
  \modelName & 8.0GB/9.3GB & 30.21s/24.85s & 0.05s/0.08s\\
  \hline
\end{tabular}}
\vspace{-0.15in}
\end{table}

\subsection{Efficiency Comparison}
For efficiency evaluation, we record the memory size, training time and prediction time. The memory size represents the required memory for training. The training time is the average time cost of an epoch. The prediction time is the time to predict all regions for one time interval. Note that the pre-training of the autoencoder introduced in \ref{sec:3.1.4} is conducted in advance, hence it does not incur additional time costs. The results are reported in Table~\ref{tab:efficiency}. We observe the following:

\noindent(1) The learning-based models for Chicago is larger than that for NYC, while the training time for Chicago is shorter. That is because compared to NYC, Chicago has fewer accidents, making it more difficult to predict but more efficient to train on. Thus, we chose a slightly larger version of all models for Chicago.

\noindent(2) \textbf{\modelName}~is more efficient than almost all learning based methods except \textbf{C-ViT}, since \textbf{C-ViT} trades performance for efficiency by re-formulating the problem as an image regression problem.

\begin{figure}
  \centering
  \includegraphics[width=0.9\linewidth]{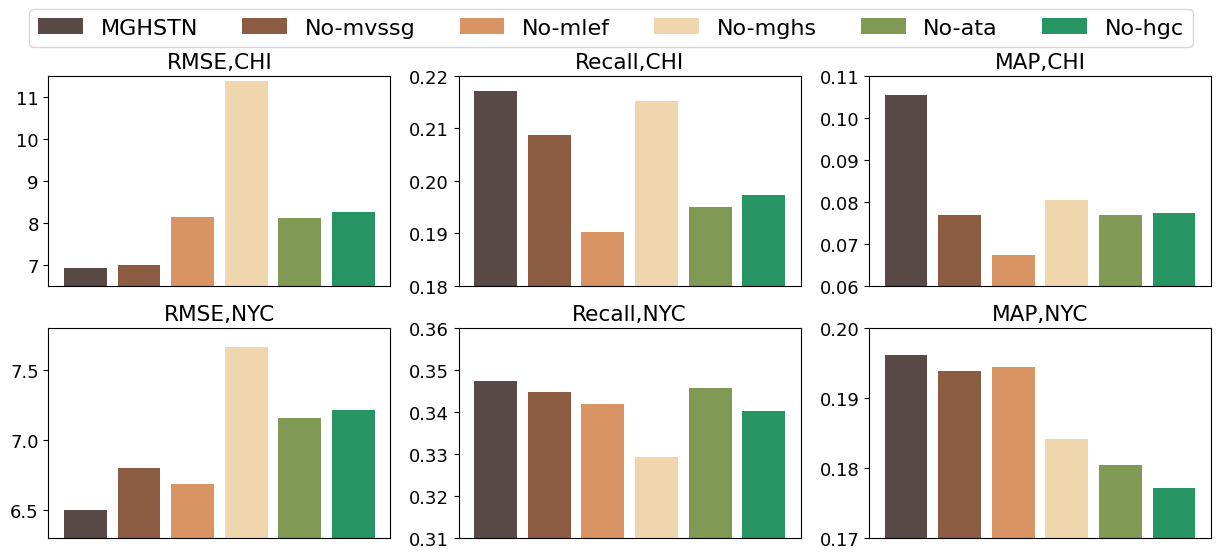}
  \vspace{-0.1in}
  \caption{Ablation study of \modelName}\label{fig:ablation}
  \vspace{-0.1in}
\end{figure}

\subsection{Ablation Study}
In this section, we perform ablation experiments to verify the effectiveness of different components. We remove multi-view semantic similarity graph, multi-level embedding fusion, multi-granularity hierarchical structure, and adaptive temporal attention module respectively as four variant models, namely \texttt{no-mvssg}, \texttt{no-mlef}, \texttt{no-mghs}, and \texttt{no-ata}. In addition, to verify the effectiveness of \textit{hierarchical graph clustering} proposed in \ref{sec:3.1.5}, we adapt uniform clustering based on spatial proximity as an alternative, namely \texttt{no-hgc}. From Fig.~\ref{fig:ablation}, we have the following observations:

\noindent(1) \textbf{\modelName}~outperforms the five variants on all metrics. In particular, the multi-granularity hierarchical structure is very critical, removing it will lead to 64.5\% performance loss on RMSE of Chicago. Meanwhile, removing the multi-level embedding fusion significantly reduce Recall and MAP, demonstrating it's ability to address the zero-inflation problem.

\noindent(2) As shown by result of \texttt{no-hgc}, replacing the \textit{hierarchical graph clustering} with uniform clustering leads to a degradation in performance. This indicates that our meticulously designed hierarchical graph clustering method is capable of effectively providing a more subtle hierarchical data structure.


\begin{figure}
  \centering
  \includegraphics[width=0.9\linewidth]{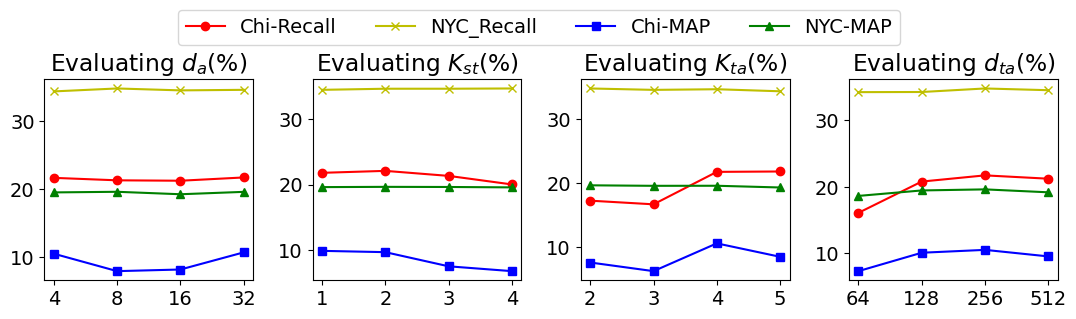}
  \vspace{-0.15in}
  \caption{Recall \& MAP vs. Hyper-parameter}\label{fig:hyper}
  \vspace{-0.1in}  
\end{figure}

\subsection{Hyper-parameter Study}
we consider the following hyper-parameters: (1) the enhancement channels $d_a$ of remote sensing enhancement; (2) the number of convolutions layers $K_{st}$ in grid convolution; (3) the number of transformer encoders $K_{ta}$ in adaptive temporal attention module; (4) the size $d_{ta}$ of feed forward in self-attention blocks. 
As shown in Fig.~\ref{fig:hyper}, we plot the Recall and MAP for different hyper-parameters. In summary, we set each hyper-parameter to the optimal: (1) For Chicago, we have $d_a=32, K_{st}=2, K_{ta}=4, d_{ta}=256$. (2) For NYC, we have $d_a=8, K_{st}=2, K_{ta}=2, d_{ta}=256$.

\section{RELATED WORK}
Traffic accident risk prediction is an important topic that has been widely studied. Related work can be mainly divided into two categories: \textit{statistics-based} and \textit{learning-based}. 
\subsection{Statistics-based Methods}
The \textit{statistics-based} methods include decision tree, SVM, k-nearest neighbor and negative binomial regression. Chong et al.~\cite{RWdectre} used decision tree and ANN model to discover new knowledge from historical data about accidents. Sharma et al.~\cite{RWsvm} leverage support vector machines (SVM) with gaussian kernel to predict traffic accident risk. The authors in~\cite{RWkneibor} identified the traffic accident potential by using the k-nearest neighbor method with real-time traffic data. Caliendo et al.~\cite{RWnbr} apply Poisson, negative binomial and negative multinomial regression models to tangents and curves respectively to predict traffic accidents. However, these works simply applied traditional statistical methods without considering the complex spatio-temporal correlation of traffic accidents and features, resulting in poor performance.

\subsection{Learning-based Methods}

In recent years, many studies have focused on using learning-based models.
For spatio-temporal prediction, many researchers employ different techniques~\cite{yuansurvey, mtmg, miao2024unified, POIjn, xu2024pefad, yuanicde, liu2022task} to model the complex spatio-temporal correlations.
Early researchers employ CNN and RNN to model spatiotemporal relationships~\cite{chen2018sdcae, hetero}. To further enhance the modeling of geographical and semantic correlations, Wang et al.~\cite{DBLP:conf/aaai/WangL0W21} introduce GSNet, which utilizes a weighted loss function to tackle the zero-inflation issue. Additionally, a multi-view multi-task model was proposed by ~\cite{MVMT}, enabling simultaneous prediction of urban fine-grained and coarse-grained traffic accident risks. Incorporating advances in computer vision, ~\cite{C-ViT} present an efficient vision transformer model, C-ViT, which processes risk map multi-channel images as input and output. Furthermore, ~\cite{hintnet, MG-TAR, uncer} utilize different network to capture traffic risk patterns and dependencies. Recently, Chen et al.~\cite{RiskContra} propose RiskContra, a contrastive learning approach with multi-kernel networks, which applies contrastive learning approach to leverages the periodic patterns to derive a tailored mixup strategy for risk sample augmentation. 
Moreover, as for related research areas like traffic prediction, multi-task learning and urban anomaly prediction, many recent studies leverage different methods to design learning-based models~\cite{stmg, STMGF, miao2022mba, mvmtmv, urban_anomaly, deepcrime, MiST}.

\section{CONCLUSION}
We studied the traffic accident risk prediction problem for urban region. We proposed a comprehensive and novel model \modelName~that is able to fully exploit spatio-temporal features by considering four significant aspects: regionality, proximity, similarity and sparsity. First, we introduced remote sensing images to capture regional background. Second, we designed two kinds of hierarchical structures for region feature and graph feature along with encoding methods tailored to them, and introduced multiple high-level prediction tasks to enhance the prediction. Third, we designed multivariate hierarchical loss function to model the problem comprehensively. Extensive experiments on real datasets verified effectiveness, efficiency and robustness of our model.

\begin{acks}
This work is supported by the National Natural Science Foundation of China under Grant No. 62032003 and Grant No. 61921003. Zhifeng Bao is supported in part by ARC DP220101434 and DP240101211.
\end{acks}

\bibliographystyle{ACM-Reference-Format}
\balance
\bibliography{ref}

\end{document}